\documentclass{iopjournal}

\usepackage{amsmath}
\usepackage{makecell}
\usepackage[style=phys]{biblatex}
\begin{document}

\articletype{Paper} 

\title{Spatiotemporal control of laser intensity using differentiable programming}

\author{Kyle G Miller$^{1,*}$\orcid{0000-0003-4826-9001}, Tomas E Gutierrez$^1$\orcid{0009-0003-7966-719X}, Archis S Joglekar$^{2,1,3}$\orcid{0000-0003-3599-5629}, Amanda Elliott$^1$\orcid{0009-0006-0020-8591}, Dustin H Froula$^{1}$\orcid{0000-0001-6981-3956}, and John P Palastro$^{1}$\orcid{0000-0002-6721-1924}}

\affil{$^1$Laboratory for Laser Energetics, University of Rochester, 250 E River Rd, Rochester, NY 14623, United States of America}

\affil{$^2$Ergodic LLC, 7511 Greenwood Avenue N \#312, Seattle, WA 98103, United States of America}

\affil{$^3$Pasteur Labs, 19 Morris Ave, Brooklyn, NY 11205, United States of America}

\affil{$^*$Author to whom any correspondence should be addressed.}

\email{kmill@lle.rochester.edu}


\begin{abstract}
Optical techniques for spatiotemporal control can produce laser pulses with custom amplitude, phase, or polarization structure. In nonlinear optics and plasma physics, the use of structured pulses typically follows a forward design approach, in which the efficacy of a known structure is analyzed for a particular application. Inverse approaches, in contrast, enable the discovery of new structures with the potential for superior performance. Here, an implementation of the unidirectional pulse propagation equation that supports automatic differentiation is combined with gradient-based optimization to design structured pulses with features that are advantageous for a range of nonlinear optical and plasma-based applications: (1)~a longitudinally uniform intensity over an extended region, (2)~a superluminal intensity peak that travels many Rayleigh ranges with constant duration, spot size, and amplitude, and (3)~a laser pulse that ionizes a gas to form a uniform column of plasma.  In the final case, optimizing the full spatiotemporal structure improves the performance by a factor of 15 compared to optimizing only spatial or only temporal structure, highlighting the advantage of spatiotemporal control.
\end{abstract}

\section{Introduction} \label{sec:intro}
Laser pulses structured in space, time, or coupled space--time provide additional degrees of freedom for optimizing laser-based applications.  The spatial structure can be two-dimensional, as in pulses with constant orbital angular momentum (OAM)~\cite{Padgett2017OrbitalInvited,Shen2019OpticalSingularities}, or three-dimensional, as in pulses with tailored trajectories~\cite{Aborahama2020DesigningDimensions} or spatially varying OAM~\cite{Aiello2015FromWheels,Dorrah2019WavelengthBeams}.
Common examples of temporal structure include pulse shape, chirp, or higher-order temporal and spectral phase.
Coupled space--time structure offers the greatest flexibility and enables custom evolution of the amplitude, phase, and polarization~\cite{Angelsky2020StructuredConcepts,Forbes2021StructuredLight,Shen2023RoadmapFields,Zhan2024SpatiotemporalTutorial}. A prominent realization is the ``flying focus,'' which features a peak intensity that moves independently of the group velocity over many Rayleigh ranges~\cite{Sainte-Marie2017,Froula2018SpatiotemporalIntensity,Jolly2020ControllingLenses,Pigeon2024UltrabroadbandPair,Liberman2024UseIntensity}.
While known structures like these present new opportunities to enhance applications~\cite{Yang2021OpticalReview,Zhu2023MultidimensionalLight,Mirhosseini2015High-dimensionalLight,Wang2022OrbitalCommunications,Caizergues2020Phase-lockedAcceleration,Palastro2020DephasinglessAcceleration,Miller2023DephasinglessRegime,Simpson2024SpatiotemporalGeneration}, they are somewhat limited in their adaptability.  The optimal structure for a particular application may combine existing structures or take an entirely new form.

The optimization of laser-based applications with structured pulses has typically followed a forward design approach, in which a known structure is evaluated for its ability to improve an outcome such as efficiency or yield. The flying focus, for instance, has found utility in plasma and nonlinear optical processes that require velocity matching or extended interaction lengths~\cite{Froula2018SpatiotemporalIntensity,Caizergues2020Phase-lockedAcceleration,Palastro2020DephasinglessAcceleration,Miller2023DephasinglessRegime,Simpson2024SpatiotemporalGeneration}. Inverse design, by contrast, begins with a desired outcome and seeks a structure that maximizes or minimizes that outcome. This approach can yield finely tuned optimizations of existing structures or reveal undiscovered structures. High-dimensional inverse design is particularly well suited to machine learning techniques for gradient-based optimization, where optimal input parameters are obtained by minimizing a loss function that is chosen to produce a desired outcome.

Advances in software tools and specialized computing architectures have made gradient-based optimization increasingly accessible to the scientific community. As an example, the JAX~\cite{jax2018github} library extends the NumPy~\cite{Harris2020ArrayNumPy} scientific library in Python by adding support for automatic differentiation (AD) and acceleration on graphical processing units (GPUs), features that can substantially speed up gradient-based optimization. This is especially true for inverse design problems that involve many inputs and few outputs, where computing finite-difference gradients is expensive~\cite{JERRELL1997AutomaticModels,Baydin2018AutomaticSurvey}.
Reverse-mode AD, in particular, efficiently computes gradients with respect to many input parameters: gradients are calculated at each step during forward evaluation and then accumulated in a reverse pass of the program---one pass for each output~\cite{Griewank2003IntroductionDifferentiation,Baydin2018AutomaticSurvey}.  This mode of AD accelerates inverse design in scientific software while adding minimal computational and programming overhead.  These advantages have motivated the adoption of AD across diverse fields, including optics~\cite{Volatier2017GeneralizationGraphs,Minkov2020InverseDifferentiation,Alagappan2023GroupNetworks,Zhang2025DynamicDifferentiation}, astronomy~\cite{Pope2021KernelDifferentiation,Desdoigts2023DifferentiableDifferentiation}, and plasma physics~\cite{McGreivy2021OptimizedDifferentiation,Joglekar2022UnsupervisedSimulations,Joglekar2023MachineSimulation,Milder2024QualitativeFusion}.

Here, we present a differentiable implementation of the unidirectional pulse propagation equation (UPPE)~\cite{Kolesik2002UnidirectionalEquation,Kolesik2004NonlinearEquations,Couairon2011PractitionersSimulation} that enables inverse design optimization of near-field spatiotemporal structure for desired features in the far field.  The optimization is performed for three examples with utility across a range of nonlinear optics and plasma physics applications. In each example, the parameters of the laser pulse in the near field are allowed to vary using a gradient descent algorithm.  The propagation of the pulse in the far field is then simulated either in vacuum or in a nonlinear medium, and the gradients are computed using AD within the JAX framework.  The three examples build in complexity: (1)~production of a static, constant intensity over an extended focal region, (2)~generation of an intensity peak that travels at a prescribed velocity while maintaining a constant spot size, duration, and amplitude, and (3)~formation of an extended, uniform region of photoionized plasma.  The first example uses pure spatial optimization, while the latter two leverage full spatiotemporal optimization.  In the final example, the desired features can only be realized with coupled space--time structure; neither spatial nor temporal structuring alone is sufficient to achieve the desired outcome. 


\section{Methods} \label{sec:methods}
\subsection{Near-field parameterization} \label{sec:new-near}
\begin{figure}
 \centering
        \includegraphics[width=0.98\textwidth]{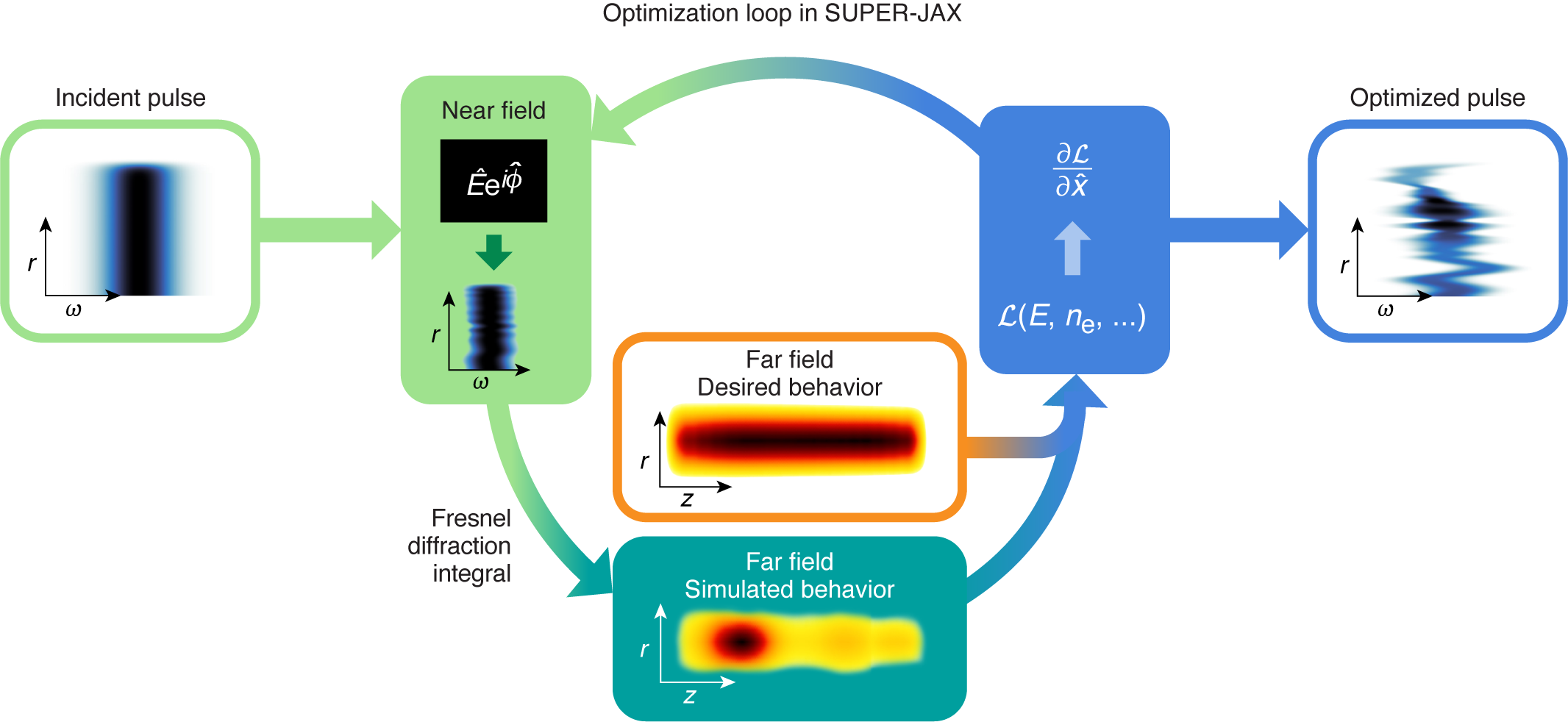}
 \caption{Diagram of the optimization loop in the differentiable program SUPER-JAX.  The spatiospectral structure of the near-field electric field $\hat{E}e^{i\hat{\phi}}$ is iteratively tuned via gradients of a loss function $\mathcal{L}$ to find an optimized pulse that produces the desired far-field behavior. Far-field propagation is simulated with the unidirectional pulse propagation equation, and gradients of $\mathcal{L}$ are calculated with automatic differentiation for each tunable parameter $\hat{x}$.}
\label{fig:schematic}
\end{figure}

Consider a laser pulse incident on an optical assembly in the near field that can freely manipulate the spatial and temporal structure of the amplitude and phase. The goal is to optimize the structure to produce desired features in the far field. The performance of the structure can be evaluated in terms of a metric, or loss function $\mathcal{L}$, that approaches a minimum as the desired features are obtained. This constitutes an inverse design problem that can be addressed with a gradient descent algorithm, where the pulse parameters are iteratively adjusted according to derivatives of the loss function with respect to those parameters (Fig.~\ref{fig:schematic}). The solution requires a framework for parameterizing the amplitude and phase that is both general enough to represent a wide variety of spatiotemporal structures and specific enough to ensure effective tuning by the optimization algorithm.

At the exit of the optical assembly, the pulse propagates in the positive $\vec{z}$ direction and is assumed to be cylindrically symmetric. In this configuration, the most general way to describe the structure of the amplitude and phase is to specify their values at every point in configuration ($r$--$t$), spectral ($k_r$--$\omega$), or spatiospectral (e.g., $r$--$\omega$) space. However, optimizing over such a large parameter space could be challenging. To facilitate optimization, reduced representations for the near-field amplitude $\hat{E}$ and phase $\hat{\phi}$ are used instead, where the hat denotes quantities that can be adjusted by the gradient descent algorithm. The amplitude is defined using familiar optical quantities, while the phase is parameterized to allow for a direct mapping onto optical elements such as diffractive optics, metasurfaces, spatial light modulators, or mirrors with custom sagitta. 

More specifically, the near-field amplitude is expressed in terms of a coefficient $\hat{A}(r)$, central frequency $\hat{\omega}(r)$, pulse duration $\hat{\tau}(r)$, and an exponent $\hat{p}$ that describes the super-Gaussian order of the spectral profile:
\begin{equation} \label{eq:e-params}
    \hat{E}(r,\omega) = \hat{A}(r) \exp \left\{ -\left( \frac{r}{w_0} \right)^{40} - \left| [\omega-\hat{\omega}(r)] \frac{\hat{\tau}(r)}{2} \right|^{\hat{p}} \right\},
\end{equation}
where $w_0$ is the near-field spot size of an approximately flat-top profile. The phase is decomposed into a polynomial expansion in frequency about a reference frequency $\omega_0$ and an expansion in cylindrically symmetric Zernike polynomials $Z_n^0(\rho)$ (defined in Appendix~\ref{sec:app-zernike}). The coefficients of this expansion define the elements of the $4 \times 4$ matrix $\hat{\mathsf{L}}$:
\begin{equation} \label{eq:phase-params}
\begin{aligned}
    \hat{\phi}(r,\omega) &= -\frac{\omega_0 r_0^2}{4cf_0} \mathsf{\Omega} \hat{\mathsf{L}} \mathsf{Z} \\
   &= -\frac{\omega_0 r_0^2}{4cf_0} \begin{pmatrix}
        1 & \frac{\Delta \omega}{\omega_0} & \left( \frac{\Delta \omega}{\omega_0} \right)^2 & \left( \frac{\Delta \omega}{\omega_0} \right)^3
    \end{pmatrix}
    \begin{pmatrix}
        \hat{l}_{00} & \hat{l}_{02} & \hat{l}_{04} & \hat{l}_{06} \\
        \hat{l}_{10} & \hat{l}_{12} & \hat{l}_{14} & \hat{l}_{16} \\
        \hat{l}_{20} & \hat{l}_{22} & \hat{l}_{24} & \hat{l}_{26} \\
        \hat{l}_{30} & \hat{l}_{32} & \hat{l}_{34} & \hat{l}_{36}
    \end{pmatrix}
    \begin{pmatrix}
        Z_0^0(r/r_0) \\
        Z_2^0(r/r_0) \\
        Z_4^0(r/r_0) \\
        Z_6^0(r/r_0)
    \end{pmatrix},
\end{aligned}
\end{equation}
where $r_0$ is the aperture radius, $c$ is the vacuum speed of light, $f_0$ is a nominal focal length, and $\Delta \omega = \omega - \omega_0$.  Element $\hat{l}_{mn}$ represents the contribution with power $m$ in frequency and radial profile $Z_n^0(r/r_0)$.  Successive powers of frequency equal to 0, 1, 2, and 3 in the matrix $\mathsf{\Omega}$ represent diffractive contributions, radial group delay, radial chirp, and radial third-order spectral phase, respectively.  An ideal lens, for example, has $\hat{l}_{00} = \hat{l}_{02} = \hat{l}_{10} = \hat{l}_{12} = 1$, with all other elements equal to zero. Once the near-field profile $\hat{E} e^{i \hat{\phi}}$ is specified, the pulse is propagated to the far field using the Fresnel diffraction integral (see Appendix~\ref{sec:app-fresnel}), providing the initial condition for far-field propagation with the UPPE.

\subsection{Far-field propagation} \label{sec:uppe}
With the initial condition determined, the UPPE is used to propagate the laser pulse through the far field. The UPPE is a reduced form of the electromagnetic wave equation that neglects backward-moving waves and their contribution to any nonlinearities~\cite{Kolesik2002UnidirectionalEquation,Kolesik2004NonlinearEquations,Couairon2011PractitionersSimulation}. By avoiding a paraxial or slowly varying envelope approximation, the UPPE is capable of modeling neutral dispersion to all orders and nonlinear processes such as harmonic generation and carrier steepening. These features have made the UPPE a standard for modeling processes such as filamentation, THz generation, and supercontinuum generation.

In the UPPE implementation used here, the cylindrically symmetric electric field is linearly polarized perpendicular to the propagation direction.
Configuration-space quantities are expressed in terms of the longitudinal coordinate $z$, radius $r$, and time $t$. Transformations to spectral space are made using quasi-discrete Hankel transforms~\cite{Guizar-Sicairos2004ComputationFields} from $r$ to the transverse wave number $k_r$ and Fourier transforms from $t$ to frequency $\omega$.  When expressed in a frame moving at speed $v_\mathrm{f}$ along the  $\vec{z}$-direction, the UPPE takes the form
\begin{equation} \label{eq:uppe}
    \frac{\partial \tilde{E}}{\partial z} - i K(k_r, \omega) \tilde{E} = \tilde{S}(k_r, \omega, \tilde{E}),
\end{equation}
where the tilde indicates a quantity is in spectral space, $\tilde{E}(k_r, \omega, z)$ is the electric field, $K(k_r, \omega) = k_z - \omega/v_\mathrm{f}$, $k_z = \sqrt{n^2(\omega) \omega^2/c^2 - k^2_r}$, and $n(\omega)$ is the linear refractive index. The right-hand side of Eq.~\eqref{eq:uppe} contains the nonlinear polarization density of the gas $\tilde{P}_\mathrm{g}$, the current density of the plasma $\tilde{J}_\mathrm{p}$, and an effective current that accounts for the energy lost due to photoionization $\tilde{J}_\mathrm{i}$:
\begin{equation} \label{eq:s}
    \tilde{S}(k_r, \omega, \tilde{E}) = \frac{\omega}{2\varepsilon_0 c^2 k_z} \left( i \omega \tilde{P}_\mathrm{g} - \tilde{J}_\mathrm{p} - \tilde{J}_\mathrm{i} \right).
\end{equation}
Each of these terms can be calculated in configuration space:
\begin{gather}
    P_\mathrm{g} = \frac{4}{3} \varepsilon_0^2 c n_2 E^3, \\
    \frac{\partial J_\mathrm{p}}{\partial t} = -\nu_\mathrm{en} J_\mathrm{p} + \frac{e^2}{m_\mathrm{e}} n_\mathrm{e} E, \label{eq:jp} \\
    J_\mathrm{i} = \frac{w(E) n_\mathrm{n} U_\mathrm{I}}{E},
\end{gather}
where $n_2$ is the nonlinear refractive index, $\nu_\mathrm{en}$ is the electron--neutral collision frequency, $n_\mathrm{e}$ is the electron density, $w(E)$ is the photoionization rate, $n_\mathrm{n}$ is the neutral density, and $U_\mathrm{I}$ is the ionization energy.  The electron density and neutral densities evolve according to $\partial_t n_\mathrm{e} = w(E) n_\mathrm{n}$ and $n_\mathrm{n} = n_0 - n_\mathrm{e}$, where $n_0$ is the initial neutral density.  In practice, Eq.~\eqref{eq:jp} is solved in spectral space to avoid integrating in time.

The UPPE is solved with a second-order predictor--corrector scheme for the nonlinear source terms (see Appendix~\ref{sec:app-solving-uppe}). The Ammosov--Delone--Krainov (ADK)~\cite{Ammosov1986TunnelField} rate is used for $w(E)$, and a fixed electron--neutral collision frequency $\nu_\mathrm{en}$ is implemented in Eq.~\eqref{eq:jp} to account for inverse bremsstrahlung absorption. The linear refractive index $n(\omega$) is obtained from the Sellmeier equation provided in Ref.~\cite{Peck1964DispersionArgon}.  All computations employ the JAX~\cite{jax2018github}, NumPy~\cite{Harris2020ArrayNumPy}, SciPy~\cite{Virtanen2020SciPyPython}, Diffrax~\cite{kidger2021on}, and Equinox~\cite{kidger2021equinox} libraries, and the complete differentiable program is called SUPER-JAX (Simulation of the Unidirectional pulse Propagation Equation at Rochester in JAX).

\subsection{Parameter implementation and optimization} \label{sec:param}
Far-field quantities, such as the electric field and electron density, are used to define a scalar loss function $\mathcal{L}$ that approaches a minimum as the quantities near the desired behavior. The minimization is accomplished with gradient-based optimization of the near-field parameters expressed by $\hat{A}(r)$, $\hat{\omega}(r)$, $\hat{\tau}(r)$, $\hat{p}$, and $\hat{\mathsf{L}}$ as defined in Sec.~\ref{sec:new-near} (and denoted by $\hat{x}$ in Fig.~\ref{fig:schematic}). The form of these parameters and their allowed variation directly influence the efficacy of the algorithm. This section describes the methodology used for the form and variation. 

Optimization algorithms are generally more stable and converge more quickly when adjustable parameters are of order unity. At the same time, the optimal values of the $\hat{l}_{mn}$ may span many orders of magnitude. To address this, each element $\hat{l}_{mn}$ in Eq.~\eqref{eq:phase-params} is implemented as
\begin{equation} \label{eq:l-params}
    \hat{l}_{mn} = l_{mn}^0 + 10^{\hat{l}_{mn}^+} - 10^{\hat{l}_{mn}^-}.
\end{equation}
The $l_{mn}^0$ term is fixed and specifies the initial phase of the electric field, e.g., 1 in the upper-left quadrant of $\hat{\mathsf{L}}$ and 0 elsewhere to model the phase applied by an ideal lens. The $\hat{l}_{mn}^+$ and $\hat{l}_{mn}^-$ terms are initialized to the same value, which sets the sensitivity of $\hat{l}_{mn}$ to variations in $\hat{l}_{mn}^{\pm}$. For example, setting $\hat{l}_{mn}^+$ and $\hat{l}_{mn}^-$ to 0 (-3) means that variations of $\pm 0.05$ in $\hat{l}_{mn}^+$ change $\hat{l}_{mn}$ by $\pm 0.1$ ($\pm 0.0001$). In other words, $l_{mn}^0$ is a constant that determines the initial phase, and $\hat{l}_{mn}^+$ and $\hat{l}_{mn}^-$ are optimized by the algorithm to find the phase $\hat{\phi}(r,\omega)$ that minimizes $\mathcal{L}$.

As opposed to the finite number of parameters that determine $\hat{\phi}$ (i.e., the 32 $\hat{l}_{mn}^{\pm}$), the coefficient, frequency, and duration comprising $\hat{E}$ should mimic continuous functions of $r$ and, in principle, may have arbitrarily many parameters. Two implementations for these functions are used here. The first method balances generality and complexity by representing $\hat{A}(r)$, $\hat{\omega}(r)$, and $\hat{\tau}(r)$ with cubic B-splines.  In this method, the parameters correspond to the weight of each B-spline, allowing the user to regulate the smoothness and variability of the functions by adjusting the number of B-splines.
To enforce physical constraints (e.g., finite bandwidth), the spline outputs are passed through a hyperbolic tangent before initializing the pulse to limit the minimum and maximum function values (see Appendix~\ref{sec:tanh}).

The second method for implementing the near-field radial functions discretizes each function into $n_\mathrm{p}$ parameters on a uniform grid along the near-field radius. In this method, the parameters are the function values on the grid. The number of parameters can be equal to or less than the number of points $n_{\mathrm{nf}}$ in the near-field radial simulation grid.  If $n_\mathrm{p} = n_{\mathrm{nf}}$, then the parameter grid exactly overlays the simulation grid, i.e., each function value is set by a single parameter. If $n_\mathrm{p} < n_{\mathrm{nf}}$, then linear interpolation is used between the parameters to compute the $n_{\mathrm{nf}}$ function values on the simulation grid.
Because this representation introduces many free parameters, the resulting functions can vary rapidly with radius.  Smoothing is applied with a moving average of 100 points on a grid containing $n_{\mathrm{nf}} = 1600$ points to reduce the variation.  Similar to the spline method, the outputs are passed through a hyperbolic tangent function to enforce constraints.

The choice of electric field parameterization strongly impacts convergence.  The three cases described in Sec.~\ref{sec:results} were evaluated with both the B-spline and discretized implementations, each of which resulted in different optimized pulses. Only the best performer is presented, with cases~1 and 3 using 20~B-splines and case~2 using 1600~discrete parameters. Other approaches to the parameterization include representing the functions with alternative basis sets or neural networks.
The convergence is also sensitive to the initial conditions, the choice of loss function, and the learning rate of the optimizer. Several optimizers were tested, and the L-BFGS-B method of the SciPy minimize function proved effective in all cases studied. Due to the many choices that affect convergence, the results presented here are not intended to be exhaustive or to represent global optima, but rather to show workable solutions that demonstrate the capabilities of AD in nonlinear optics and plasma physics.

\section{Results} \label{sec:results}
This section presents three simulations that demonstrate the utility and flexibility of gradient-based optimization for the inverse design of structured laser pulses.  The first two cases are simulated in vacuum, while the third includes the effects of all nonlinear source terms in Eq.~\eqref{eq:s}.  Case~1 is initialized with a monochromatic, continuous wave and produces an extended focal region of constant intensity.  Case~2 starts with a finite-duration pulse and creates a superluminal intensity peak that travels many Rayleigh ranges with constant spot size, duration, and amplitude.  Case~3 modifies a finite-duration pulse to generate a uniform column of plasma.  The initial phase of the electric field is always set to that of an ideal focusing lens.  In each case, the loss function $\mathcal{L}$ is reduced by at least a factor of 15 from its initial value $\mathcal{L}_0$.

Table~\ref{tab:setup} summarizes the parameters for each simulation, including the near-field pulse profile, far-field domain, and optimization metrics.  The initial wavelength $\lambda_0$ corresponds to the frequency $\omega_0$ at which $\hat{\omega}$ is initialized, and $U$ is the initial pulse energy.  Physical constraints on $\hat{A}$, $\hat{\omega}$, and $\hat{\tau}$ are listed as minimum, initial, and maximum values (e.g., $A_\mathrm{min}$, $A_0$, and $A_\mathrm{max}$).  The number of UPPE grid points are given in radius, time, and longitudinal space as $n_r$, $n_t$, and $n_z$, respectively.  The corresponding physical dimensions of the domain are the maximum radius $R_\mathrm{max}$, temporal width $T$, and longitudinal width $L_z$.  The longitudinal distance over which the loss function is integrated is given as $L_\mathrm{loss}$.  The number of iterations $n_\mathrm{opt}$ and the total wall-clock time $T_\mathrm{wall}$ are also included. For case~1, convergence was reached after 409 iterations (1.7~hrs), whereas cases~2 and 3 continued optimizing for the full job duration (maximum of 48~hrs).

\begin{table}
\caption{Simulation parameters for the three cases presented in Sec.~\ref{sec:results}.  Parameters are given for the near-field pulse profile, the far-field domain, and the optimization algorithm.}
\centering
{\small
\begin{tabular}{l | c | c | c}
\hline
& \makecell{Case~1: longitudinally \\ uniform intensity} & \makecell{Case~2: spatiotemporally \\ uniform intensity} & \makecell{Case~3: spatially \\ uniform density} \\
\hline
$\lambda_0$ ($\mu$m) & 1 & 1 & 1 \\
$U$ (mJ) & -- & 0.1 & 0.3 \\
$w_0$ (cm) & 3 & 2.85 & 3.64 \\
$r_0$ (cm) & 3 & 3 & 3.64 \\
$f_0$ (cm) & 63 & 63 & 51 \\
$n_{\mathrm{nf}}$ & 1600 & 1600 & 1600 \\
$\hat{A}$, $\hat{\tau}$, $\hat{\omega}$ form & 20 B-splines & 1600 discrete points & 20 B-splines \\
$A_\mathrm{min}$, $A_0$, $A_\mathrm{max}$ & 0.25, 1, 1.75 & 0.25, 1, 1.75 & 0.25, 1, 1.75 \\
$\omega_\mathrm{min}$, $\omega_0$, $\omega_\mathrm{max}$ ($\omega_0$) & -- & 0.9, 1, 1.1 & 0.99, 1, 1.01 \\
$\tau_\mathrm{min}$, $\tau_0$, $\tau_\mathrm{max}$ (fs) & -- & 14, 17.5, 1000 & 1261, 1375, 1489 \\
$\hat{p}$ & -- & 3.63 & 0.93 \\
$\hat{\mathsf{L}}$ variation & $\hat{l}_{0n}$ & $\hat{l}_{mn}$ & $\hat{l}_{mn}$ \\
\hline
$n_r$ & 2000 & 1000 & 622 \\
$n_t$ & 6 & 768 & 16384 \\
$n_z$ & 300 & 200 & 800 \\
$R_\mathrm{max}$ (mm) & 2 & 1 & 0.622 \\
$T$ (fs) & 2.8 & 600 & 9500 \\
$L_z$ (cm) & 3 & 1.6 & 0.32 \\
\hline
$L_\mathrm{loss}$ (cm) & 1.7 & 0.8 & 0.16 \\
$n_\mathrm{opt}$ & 409 & 3916 & 1161 \\ 
$T_\mathrm{wall}$ (hr) & 1.7 & 48 & 48 \\
$\mathcal{L} / \mathcal{L}_0$ & 0.015 & 0.023 & 0.066 \\
\hline
\end{tabular}
}
\label{tab:setup}
\end{table}

\subsection{Case~1: longitudinally uniform intensity} \label{sec:linear-1}
Unlike ideal lenses, which focus laser beams to a single plane, axicons~\cite{Sochacki1992NonparaxialAxicons,Friberg1996Stationary-phaseAxicons} and axiparabolas~\cite{Smartsev2019Axiparabola:Lasers} produce longitudinally extended focal regions.  Although these optics can be designed to yield an approximately uniform intensity profile, they often create longitudinal modulations in the on-axis intensity. Many applications that rely on an extended focal region would benefit from a truly constant intensity profile.

\begin{figure}
 \centering
        \includegraphics[width=0.75\textwidth]{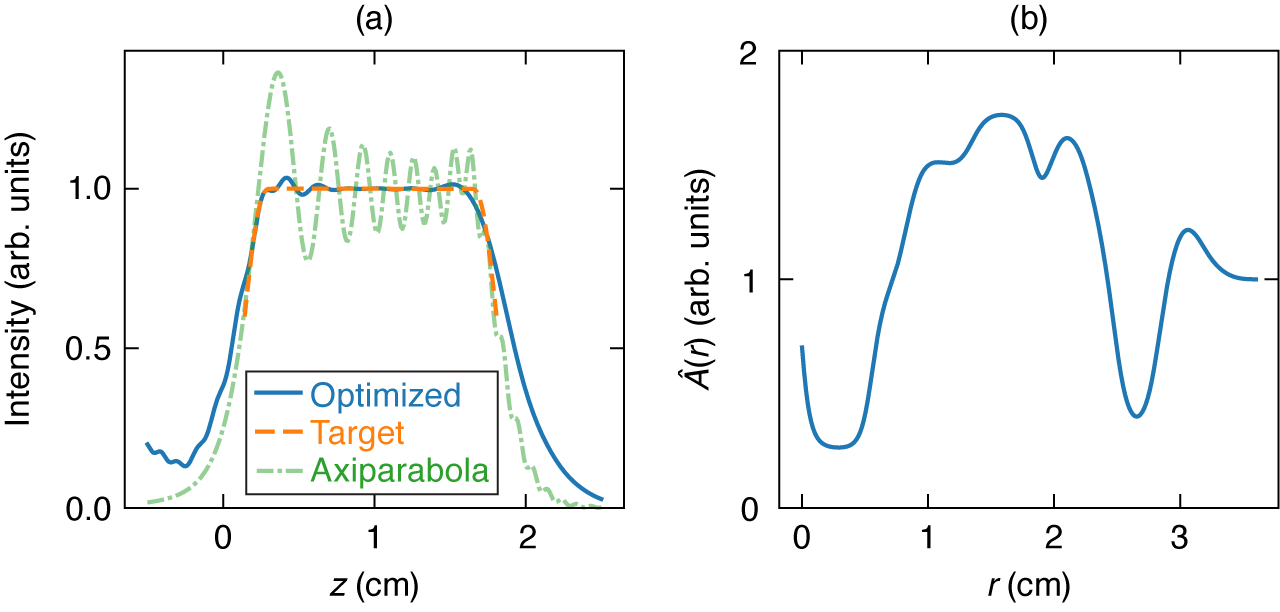}
 \caption{Optimization for case~1: producing an extended, uniform focal region with a monochromatic laser beam.  (a)~The simulated (solid blue) and target (dashed orange) on-axis intensity of the laser beam.  For comparison, the on-axis intensity produced by an axiparabola with a 2-cm focal region is also shown (dashed--dotted green).  (b)~The near-field coefficient $\hat{A}$ as a function of radius, parameterized with 20~B-splines.}
\label{fig:linear-1}
\end{figure}

The goal of the first simulation is to produce a longitudinally extended focal region with uniform intensity in vacuum.  A monochromatic electric field is considered, requiring only spatial structuring through $\hat{A}$ and the first row of $\hat{\mathsf{L}}$ (corresponding to zeroth-order terms in powers of frequency).  The loss function is defined as $\mathcal{L} = \int |I_\mathrm{sim}(r=0,z) - I_\mathrm{target}(z)| \,\mathrm{d}z$, where $I = c \varepsilon_0|E|^2 / 2$ is the intensity.  Figure~\ref{fig:linear-1} shows (a)~the optimized on-axis intensity profile (blue) alongside the target intensity profile (dashed orange), and (b)~the optimal near-field coefficient $\hat{A}$.  The converged intensity is relatively flat and closely matches the target across the region included in the loss function.  The optimal coefficients from the first row of $\hat{\mathsf{L}}$ are listed in the first row of Tab.~\ref{tab:lin-1}.

\begin{table}
\caption{The learned $\hat{l}_{0n}$ coefficients for case~1, corresponding to row ``$\mathsf{\Omega}_0$, learned'' and column $Z_n^0$.  For comparison, the Zernike coefficients of an axiparabola with a 2-cm positive focal region are shown in the row ``$\mathsf{\Omega}_0$, axiparabola.''  This is the same axiparabola used to produce the green dashed--dotted profile in Fig.~\ref{fig:linear-1}(a).}
\centering
{\small
\begin{tabular}{l c c c c}
\hline
& $Z_0^0$ & $Z_2^0$ & $Z_4^0$ & $Z_6^0$ \\
\hline
$\mathsf{\Omega}_0$, learned & 1.0000 & 0.9918 & 0.0082 & -0.0005 \\
\hline
$\mathsf{\Omega}_0$, axiparabola & 1.0000 & 0.9844 & -0.0051 & 0.0000 \\
\hline
\end{tabular}
}
\label{tab:lin-1}
\end{table}

An axiparabola generates an extended focal region that resembles the target intensity profile, albeit with modulations in the on-axis intensity.  The green dashed--dotted line in Fig.~\ref{fig:linear-1}(a) shows the intensity pattern for an axiparabola with a positive focal region of 2~cm, specified by the Zernike coefficients listed in the second row of Tab.~\ref{tab:lin-1}. These coefficients are similar to those learned by the algorithm, but the $\hat{l}_{04}$ term has opposite sign. The additional element $\hat{l}_{06}=-0.0005$ in the learned phase is an important modification, without which the loss value rises by a factor of 6. For comparison, the axiparabola with a radially uniform $\hat{A}$ results in a loss value $7\times$ greater than the optimum.  This comparison illustrates how forward design can identify pulse-shaping techniques that produce a desirable effect, while inverse design can further refine or discover new pulse structures for specific applications.

\subsection{Case~2: spatiotemporally uniform intensity} \label{sec:linear-2}
Many applications---such as laser wakefield acceleration (LWFA)~\cite{Caizergues2020Phase-lockedAcceleration,Palastro2020DephasinglessAcceleration,Miller2023DephasinglessRegime} and THz generation~\cite{Simpson2024SpatiotemporalGeneration}---require an intensity peak that moves with a prescribed velocity, finite duration, and constant spot size over an extended region.  Several methods have been proposed to produce such intensity peaks~\cite{Sainte-Marie2017,Froula2018SpatiotemporalIntensity,Jolly2020ControllingLenses,Simpson2022SpatiotemporalModulation,Ambat2023Programmable-trajectoryPulses}, but these approaches involve trade-offs among non-uniform intensity, variable spot size, and limitations on the minimum pulse duration.  As an example, an axiparabola combined with a radially stepped echelon produces an intensity peak with an ultrashort duration and a custom trajectory favorable for LWFA~\cite{Palastro2020DephasinglessAcceleration,Ambat2023Programmable-trajectoryPulses}, but the resulting spot size variation can complicate the interaction~\cite{Miller2023DephasinglessRegime}.

The goal of the second simulation is to create an intensity peak with uniform intensity, duration (40~fs), and spot size (10~$\mu$m) that propagates at a superluminal velocity ($1.005c$) over many Rayleigh ranges (8~mm, or 27~Rayleigh ranges).  All parameters in $\hat{E}$ and $\hat{\phi}$ are allowed to vary, and $\hat{A}$, $\hat{\tau}$, and $\hat{\omega}$ are implemented with 1600 discrete points.  The pulse energy is held constant at 0.1~mJ.  The loss function is defined as $\mathcal{L} = \int |I_\mathrm{sim}(r,t,z) - I_\mathrm{target}(r,t,z)| \, \mathrm{d}r \, \mathrm{d}t \, \mathrm{d}z$, and regions in $r$ and $t$ are included only where $I_\mathrm{target}$ is within 90\% of its maximum value.

\begin{figure}
 \centering
        \includegraphics[width=0.95\textwidth]{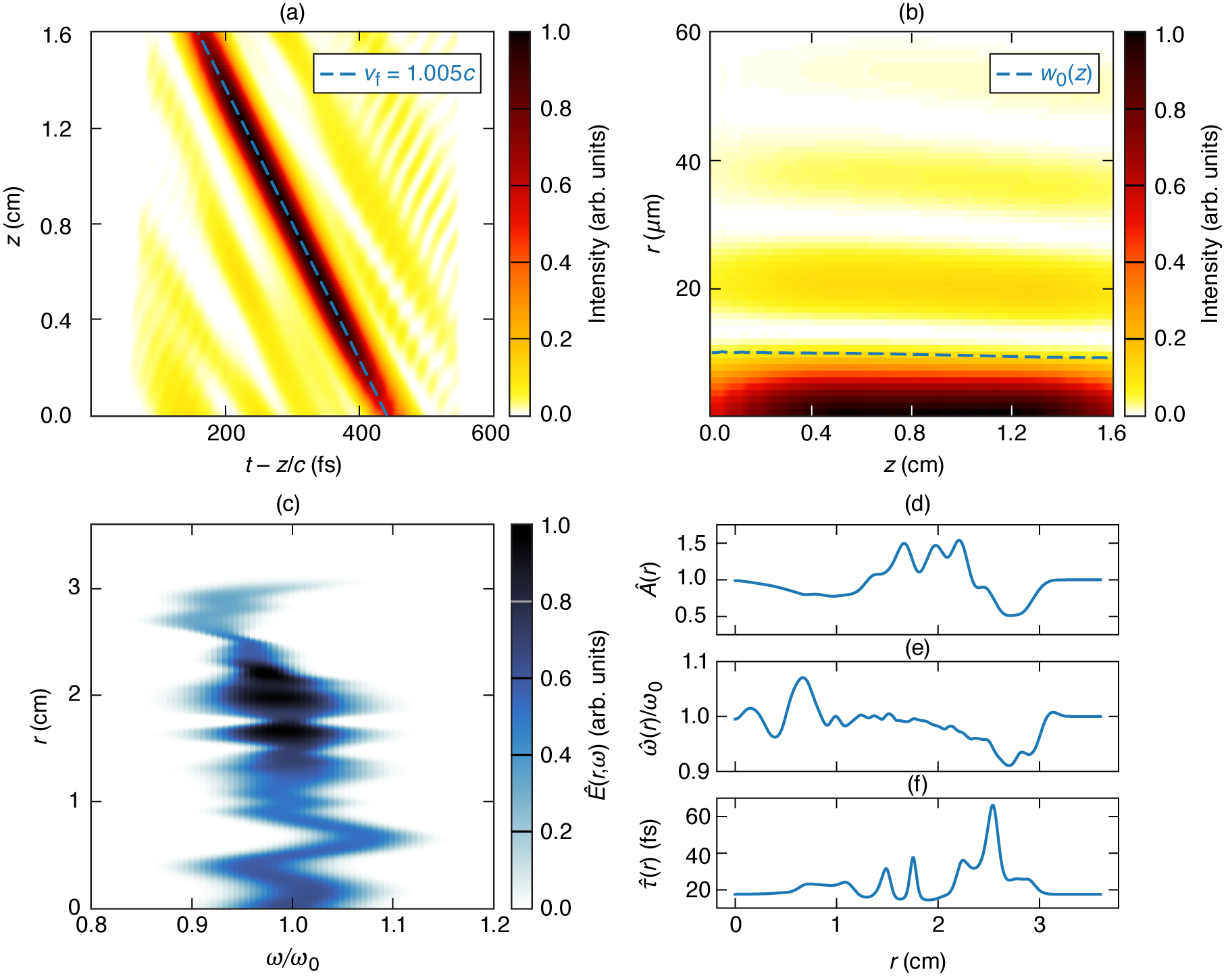}
 \caption{Optimization for case~2: creating a uniform, superluminal ($1.005c$) intensity peak with constant duration (40~fs) and spot size (10~$\mu$m).  (a)~The optimized intensity profile as a function of $z$ and $t-z/c$, which exhibits a focal velocity of $1.005c$ (dashed blue) over 8~mm.  (b)~Radial profile at the temporal location of maximum intensity along $z$.  The fitted spot size (dashed blue) varies by only 6\% over the target region.  (c)~Near-field electric field spectral amplitude.  The individual near-field functions are shown for (d)~$\hat{A}$, (e)~$\hat{\omega}$, and (f)~$\hat{\tau}$.}
\label{fig:linear-2}
\end{figure}

The optimized intensity profile closely reproduces the desired behavior and is shown in Fig.~\ref{fig:linear-2} as a function of (a)~$z$ and $t-z/c$ and (b)~$r$ and $z$.  In (a), the intensity peak propagates with the specified superluminal velocity, and regions of high intensity are mainly concentrated along the target trajectory (dashed line).
The spot size in (b) is fit to $1/e^2$ of the intensity and remains nearly constant, varying by only 6\% over the region of interest (between 0.4 and 1.2~mm in $z$).  For comparison, an axiparabola--echelon combination that achieves the same focal velocity, focal region, and duration has a spot size that varies by a factor of 2.4 over the target region~\cite{Smartsev2019Axiparabola:Lasers,Ambat2023Programmable-trajectoryPulses,Miller2023DephasinglessRegime}.  The near-field pulse shape exhibits variations in the coefficient, central frequency, and duration [see Fig.~\ref{fig:linear-2}(c--f)], yielding a complex structure that is unlikely to result from forward design.  The coefficient function shown in (d) is largest between approximately 1.4 and 2.3~cm.  This radial concentration of the pulse energy was also employed in Ref.~\cite{Friberg1996Stationary-phaseAxicons} to reduce spot size variation when focusing with axicons.

\begin{table}
\caption{The learned $\hat{l}_{mn}$ coefficients for case~2, corresponding to row $\mathsf{\Omega}_m$ and column $Z_n^0$.}
\centering
{\small
\begin{tabular}{l c c c c}
\hline
& $Z_0^0$ & $Z_2^0$ & $Z_4^0$ & $Z_6^0$ \\
\hline
$\mathsf{\Omega}_0$ & 1.0000 & 1.0158 & 0.0003 & -0.0017 \\
$\mathsf{\Omega}_1$ & 1.0000 & 1.2087 & 0.0014 & -0.0001 \\
$\mathsf{\Omega}_2$ & -0.9872 & 0.0049 & 0.0382 & 0.0000 \\
$\mathsf{\Omega}_3$ & 0.0001 & -0.0002 & 0.0000 & 0.0000 \\
\hline
\end{tabular}
}
\label{tab:lin-2}
\end{table}

All 16 coefficients of $\hat{\mathsf{L}}$ are included in the optimization of this dynamic profile, and their values in Tab.~\ref{tab:lin-2} reveal a richer structure than in the static case (Tab.~\ref{tab:lin-1}).  Notably, a large chirp is applied via the coefficient $\hat{l}_{20}=-0.9872$.  Chirped pulses have been employed in previous flying-focus implementations~\cite{Sainte-Marie2017,Froula2018SpatiotemporalIntensity,Jolly2020ControllingLenses}, but typically with diffractive optics (e.g., $\hat{l}_{00} = \hat{l}_{02} = 1$ with all other elements of $\hat{\mathsf{L}}$ set to 0).  Here, the chirp is combined with a chromatic focusing optic exhibiting additional spherical aberration. This aberration plays a key role; for instance, without the large $\hat{l}_{24}$ term the loss value increases by a factor of 41.  The third-order spectral phase terms in the last row of Tab.~\ref{tab:lin-2} remain small.

\subsection{Case~3: Spatially uniform density} \label{sec:nonlinear-2}
Cases~1 and 2 focused on producing an extended, uniform intensity in the far field.  Many plasma-based applications further benefit from a uniform, pre-ionized plasma density that enables pulses to propagate without extraneous energy loss or distortion from ionization effects. Localized ionization can cause refraction, and plasma introduces additional dispersion during propagation.  Because of these effects, the optimal structured pulse for creating a uniform plasma column may differ from a structured pulse that produces a uniform intensity in vacuum.

\begin{figure}
 \centering
        \includegraphics[width=0.95\textwidth]{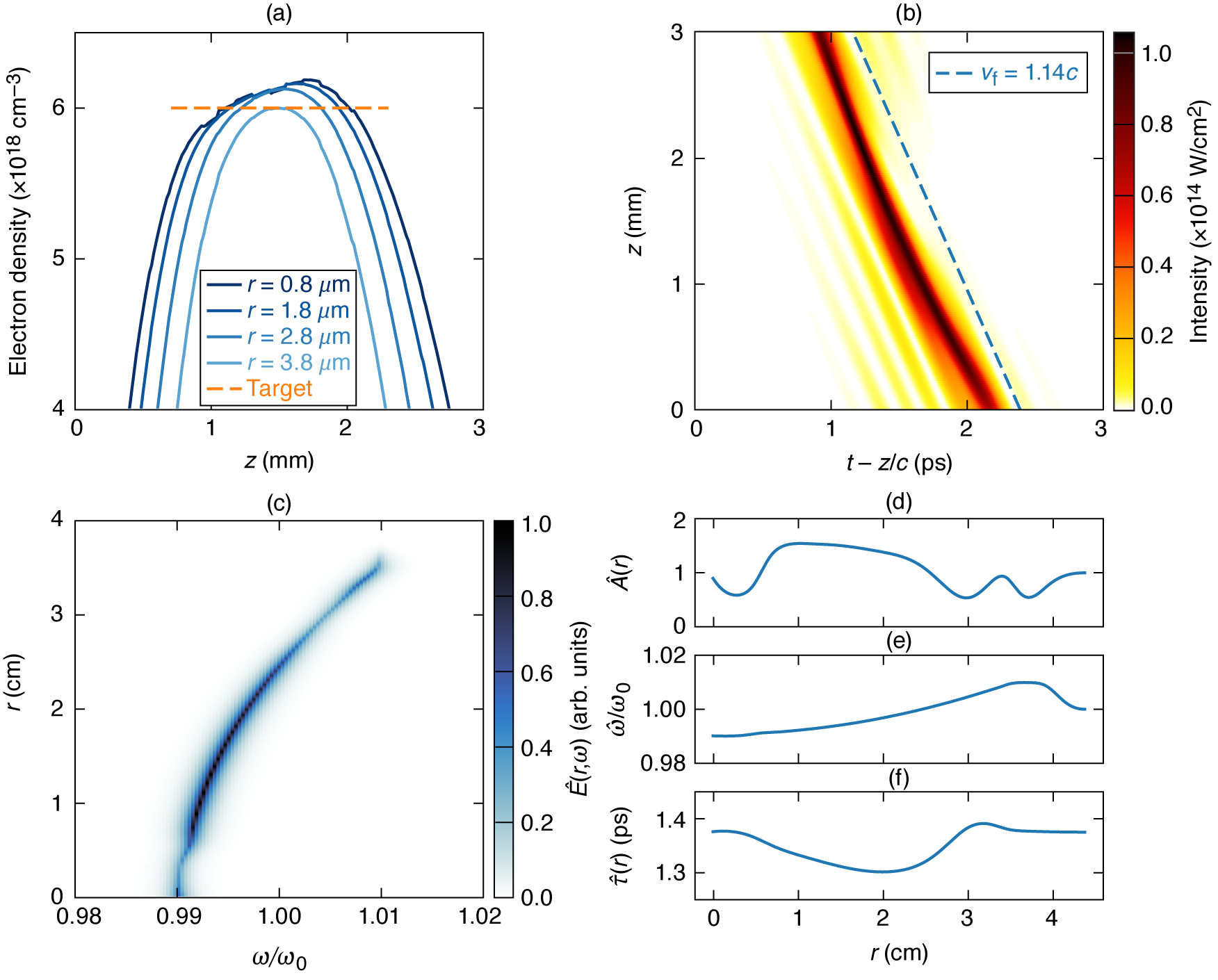}
 \caption{Optimization for case~3: producing a uniform-density plasma column.  (a)~Lineouts of the electron density at the final simulation time for the four innermost radial cells (solid blue), compared with the target density (dashed orange).  (b)~The on-axis far-field intensity profile, which propagates at a superluminal velocity of 1.14$c$.  (c)~Near-field electric field spectral amplitude of the optimized pulse, which exhibits a central frequency that increases quadratically with radius.  The individual near-field functions for (d)~$\hat{A}$, (e)~$\hat{\omega}$, and (f)~$\hat{\tau}$ are also shown, each implemented with 20~B-splines.}
\label{fig:nonlinear-2}
\end{figure}

\begin{table}
\caption{The learned $\hat{l}_{mn}$ coefficients for case~3, corresponding to row $\mathsf{\Omega}_m$ and column $Z_n^0$.}
\centering
{\small
\begin{tabular}{l c c c c}
\hline
& $Z_0^0$ & $Z_2^0$ & $Z_4^0$ & $Z_6^0$ \\
\hline
$\mathsf{\Omega}_0$ & 1.0000 & 0.9773 & -0.0001 & 0.0001 \\
$\mathsf{\Omega}_1$ & 1.0000 & 0.9889 & 0.0000 & 0.0000 \\
$\mathsf{\Omega}_2$ & -0.0002 & 0.0000 & 0.0000 & 0.0000 \\
$\mathsf{\Omega}_3$ & 0.0000 & 0.0000 & 0.0000 & 0.0000 \\
\hline
\end{tabular}
}
\label{tab:non-2}
\end{table}

The third and final simulation aims to generate a 1.6-mm-long, 4-$\mu$m-radius column of uniform plasma using a single laser pulse. Hydrogen gas at atmospheric density is used as the propagation medium, with an ionization potential $U_\mathrm{I}=13.6$~eV and a nonlinear refractive index $n_2 = 7.5\times 10^{-24}$~m$^2$/W. The pulse energy is held constant at 0.3~mJ, but all other parameters are allowed to vary. The near-field radial functions are implemented with 20~B-splines. The loss function, evaluated at the final simulation time $t_\mathrm{f}$, is defined as $\mathcal{L} = \int |n_\mathrm{sim}(r,t_\mathrm{f},z) - n_\mathrm{target}| \, \mathrm{d}r \, \mathrm{d}z$, where the integration is restricted to the central 1.6~mm in $z$ and the four innermost radial cells ($r < 4$~$\mu$m).

The converged results shown in Fig.~\ref{fig:nonlinear-2} demonstrate a type of superluminal flying focus similar to the ``flying focus X,'' where the central frequency increases quadratically with radius~\cite{Simpson2022SpatiotemporalModulation,Ramsey2023ExactFocus}.  The electron density is shown in (a) as a function of $z$ for the four innermost radial cells (shades of blue) along  with the target density of $6\times10^{18}$~cm$^{-3}$ (dashed orange).  The electron density profile produced by the pulse is nearly equal to the target value across most of the region.  The (b)~far-field on-axis intensity peak propagates with a superluminal velocity of 1.14$c$, and the (c)~near-field electric field amplitude has a central frequency that increases quadratically with radius.  The individual near-field coefficient, central frequency, and duration are plotted in (d--f).  All 16 coefficients of $\hat{\mathsf{L}}$ shown in Tab.~\ref{tab:non-2} were allowed to vary, yet the values indicate only modest chromatic focusing, primarily through the $\hat{l}_{02}$ and $\hat{l}_{12}$ terms.  This slight deviation from an ideal lens, combined with the spectral amplitude shown in Fig.~\ref{fig:nonlinear-2}(c), is sufficient to produce the target density.

\begin{figure}
 \centering
        \includegraphics[width=0.95\textwidth]{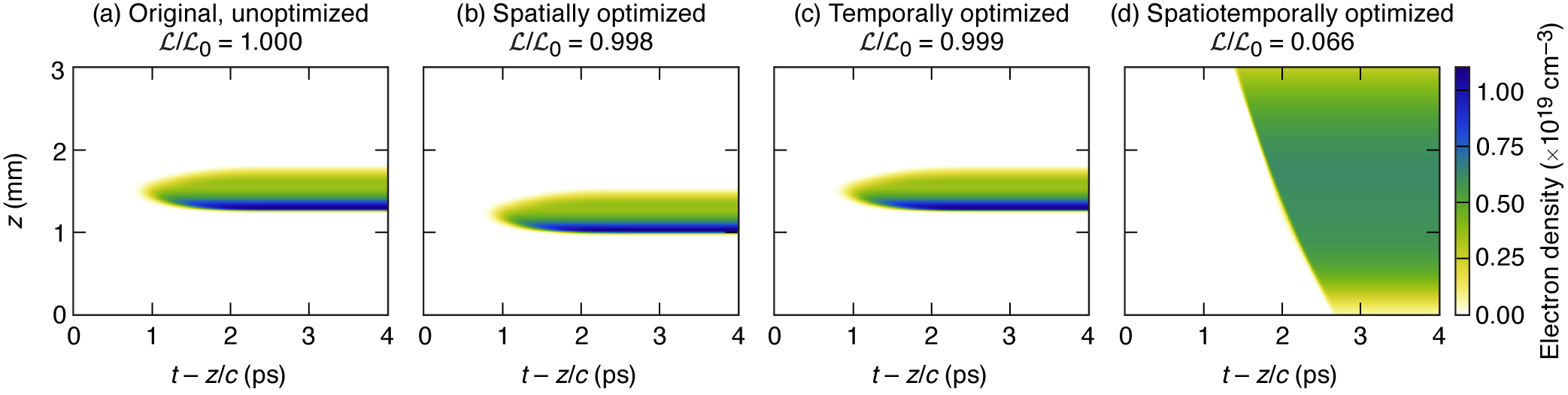}
 \caption{On-axis density profiles as a function of $z$ and $t-z/c$ for the labeled cases, along with the ratio of final to initial loss values $\mathcal{L}/\mathcal{L}_0$.  (a)~A laser pulse focused to the nominal focal point with no optimization.  Cases optimized with (b)~only spatial and (c)~only temporal structuring are also shown, for which there is no appreciable reduction in the loss value.  (d)~Full optimization (same simulation shown in Fig.~\ref{fig:nonlinear-2}), for which the loss value is reduced by 93\%.}
\label{fig:nonlinear-2-vary}
\end{figure}

Two additional tests were performed to demonstrate the importance of spatiotemporal structure in creating a uniform plasma profile: the first optimized only spatial structure (i.e., $\hat{A}$ and the first two rows of $\hat{\mathsf{L}}$, with $\hat{l}_{0n} = \hat{l}_{1n}$), and the second optimized only temporal structure (i.e., $\hat{\omega}$ and $\hat{\tau}$ as constants rather than radial functions, $\hat{p}$, and the first column of $\hat{\mathsf{L}}$, with $\hat{l}_{02} = \hat{l}_{00}$ and $\hat{l}_{12} = \hat{l}_{10}$).  The resulting density profiles are shown in Fig.~\ref{fig:nonlinear-2-vary}(b) and (c), respectively, along with (a)~the original, unoptimized case and (d)~the optimized case with full spatiotemporal structuring.  The ratio of final to initial loss values $\mathcal{L}/\mathcal{L}_0$ is reported for each case.  Optimizing with only spatial and only temporal structuring in (b) and (c) gives almost no improvement, reducing the loss by just 0.2\% and 0.1\%, respectively.  A large improvement is achieved only when the phase and amplitude are optimized together in (d), reducing the loss by 93\%.

\section{Discussion and conclusion} \label{sec:conclusion}
A differentiable implementation of the UPPE enables the inverse design of spatiotemporal structures optimized for both linear and nonlinear processes. The optimization loop involves four steps. First, the near-field spatiospectral amplitude and phase of an incident laser pulse are parameterized as described in Sec.~\ref{sec:new-near}. The pulse is then propagated to the far field, after which the UPPE system presented in Sec.~\ref{sec:uppe} is solved using the JAX Python package---complete with native GPU acceleration and AD. The third step evaluates a customized single-output loss function that is defined to yield desired features in the far field when minimized. Finally, the near-field parameters are optimized by a gradient descent algorithm, with gradients of the loss function obtained through reverse-mode AD.

Optimization was performed for the three test cases described in Sec.~\ref{sec:results}, and the loss function decreased substantially in each case (see the last row of Tab.~\ref{tab:setup}). In case~1, optimization of the coefficient $\hat{A}$ and the phase demonstrated the ability of inverse design to improve upon existing structures, such as those produced by an axiparabola.  In case~2, the full $\hat{E}e^{i\hat{\phi}}$ expression was optimized to produce a rich near-field structure that would likely only emerge through inverse design. The optimization in case~3 generated a type of flying focus that achieved the desired behavior, which was only possible with full spatiotemporal control.

The utility of the differentiable framework ultimately hinges on whether the structured pulses can be realized in the laboratory.  For instance, due to its few-percent bandwidth and relatively smooth profile, the near-field electric field profile obtained in case~3 [Fig.~\ref{fig:nonlinear-2}(a)] may be easier to realize than the profile from case~2 [Fig.~\ref{fig:linear-2}(c)].  Further studies could investigate the significance of the rapid variations in $\hat{A}$, $\hat{\omega}$, and $\hat{\tau}$ observed in Fig.~\ref{fig:linear-2}(d--f).  These functions could be smoothed, then simulated again to test whether the variations are essential for good performance.  Alternatively, the optimization loops could be restarted using the converged solution (or a smoothed version of it) as an initial condition.  Restarting the loop in this way could improve the converged result, since the L-BFGS-B minimization method retains a history of gradients and parameter updates.  A stochastic gradient descent algorithm could also be tested.  Once a sufficiently simple pulse structure is determined for a given application, a range of experimental techniques for generating spatiotemporally structured light~\cite{Chen2018AVisible,Shaltout2019SpatiotemporalMetasurfaces,Divitt2019UltrafastMetasurfaces,Simpson2022SpatiotemporalModulation,Chen2022SynthesizingControl,Panuski2022AModulator,Franco2023Curve-shapedBehavior,Ambat2023Programmable-trajectoryPulses} could then be leveraged to produce the desired pulse.

The parameterization of the incident electric field amplitude and phase described in Sec.~\ref{sec:new-near} is only one possible approach to characterizing spatiotemporal pulse shapes, and other parameterizations may yield pulses with improved performance.  Several descriptions more general than Eqs.~\eqref{eq:e-params} and \eqref{eq:phase-params} were tested: using B-splines to represent a functional form $\hat{F}[\left(\omega - \hat{\omega}(r) \right) * \hat{\tau}(r)]$ for the spectral amplitude instead of a super-Gaussian, a two-dimensional B-spline representation of $\hat{E}(r, \omega)$, and replacing the $\hat{\mathsf{L}} \mathsf{Z}$ matrix product with four free-form functions of radius $\hat{l}_m(r)$.  In each case, however, the optimization algorithm produced noisy far-field solutions with larger final loss values compared to the results presented in Sec.~\ref{sec:results}.  The right balance must be found between generality and tractability in choosing the field parameterization.

As tools for AD and GPU acceleration become more widely available, gradient-based optimization of physics simulations is poised to become an important tool for research and discovery.  For future work, this differentiable framework could be applied to the inverse design of structured pulses for other nonlinear optical processes, including perturbative harmonic generation, high-harmonic generation, long-distance pulse propagation in air sustained by self-focusing and ionization refraction, and the generation of terahertz radiation.  Further analysis of the results presented here could yield deeper insight into the spatiotemporal structure of the optimized pulses and guide strategies for their experimental realization.

%
%


\funding{This report was prepared as an account of work sponsored by an agency of the U.S. Government. Neither the U.S. Government nor any agency thereof, nor any of their employees, makes any warranty, express or implied, or assumes any legal liability or responsibility for the accuracy, completeness, or usefulness of any information, apparatus, product, or process disclosed, or represents that its use would not infringe privately owned rights. Reference herein to any specific commercial product, process, or service by trade name, trademark, manufacturer, or otherwise does not necessarily constitute or imply its endorsement, recommendation, or favoring by the U.S. Government or any agency thereof. The views and opinions of authors expressed herein do not necessarily state or reflect those of the U.S. Government or any agency thereof.

This material is based upon work supported by the Office of Fusion Energy Sciences under Award Numbers DE-SC0021057, the Department of Energy (DOE) [National Nuclear Security Administration (NNSA)] University of Rochester “National Inertial Confinement Fusion Program” under Award Number DE-NA0004144, and the New York State Energy Research and Development Authority. Simulations were performed at NERSC under m4372.}

\roles{K.G.M. was involved in Conceptualization, Data curation, Formal analysis, Investigation, Methodology, Resources, Software, Validation, Visualization, Writing – original draft, and Writing – review \& editing.
T.E.G. was involved in Formal analysis, Investigation, Software, Validation, and Writing – review \& editing.
A.S.J. was involved in Conceptualization, Data curation, Methodology, Resources, Software, and Writing – review \& editing.
A.E. was involved in Software.
D.H.F. was involved in Conceptualization, Funding acquisition, Project administration, Resources, and Writing – review \& editing.
J.P.P. was involved in Conceptualization, Methodology, Project administration, Resources, Software, Supervision, Validation, Visualization, Writing – original draft, and Writing – review \& editing.}

\data{The datasets generated and/or analyzed during the current study and the software used to generate these datasets are available from the corresponding author on reasonable request.}

\appendix

\makeatletter
\renewcommand\@seccntformat[1]{Appendix \csname the#1\endcsname.\quad}
\makeatother

\section{Zernike polynomial definitions} \label{sec:app-zernike}
The normalization for the Zernike polynomials is defined such that $(2n+2)\int_0^1 Z_n^0(\rho) Z_{n'}^0(\rho) \rho \, \mathrm{d}\rho = \delta_{nn'}$.  The four Zernike polynomials used in Eq.~\eqref{eq:phase-params} are then
\begin{align}
    Z_0^0(\rho) &= 1 \\
    Z_2^0(\rho) &= 2\rho^2 - 1 \\
    Z_4^0(\rho) &= 6\rho^4 - 6\rho^2 + 1 \\
    Z_6^0(\rho) &= 20\rho^6 - 30\rho^4 + 12\rho^2 - 1.
\end{align}

\section{Transforming from the near field to the far field} \label{sec:app-fresnel}
The near-field electric field $\hat{E}(r,\omega) e^{i \hat{\phi}(r,\omega)}$ is computed from the product of the spectral amplitude and phase, which are defined in Eqs.~\eqref{eq:e-params} and \eqref{eq:phase-params}, respectively.  The far-field electric field at focus is found by computing the Fresnel diffraction integral as follows:
\begin{equation}
    \tilde{E}(r, \omega, f_0) = -\frac{i \omega }{cf_0}e^{\frac{i\omega r^2}{2cf_0}} \int \hat{E}(r', \omega) e^{i\hat{\phi}(r', \omega)} e^{\frac{i \omega r'^2}{2cf_0}} J_0 \left( \frac{\omega r r'}{cf_0} \right) r' \, \mathrm{d} r',
\end{equation}
where $r'$ is the near-field radius and $J_0$ is the zeroth-order Bessel function of the first kind. To initialize the simulation at the specified longitudinal location $z_0$, the transformed electric field is multiplied by the linear evolution operator,
\begin{equation}
    \tilde{E}(k_r, \omega, z_0) = \tilde{E}(k_r, \omega, f_0) e^{iK(z_0-f_0)},
\end{equation}
which can be derived from Eq.~\eqref{eq:app-estar} with $\tilde{S}=0$.

\section{Predictor--corrector scheme for the unidirectional pulse propagation equation} \label{sec:app-solving-uppe}
This appendix details the predictor--corrector scheme used to solve the UPPE in the spectral domain, as defined in Eq.~\eqref{eq:uppe}. First, the electric field is written as $\tilde{E} = \tilde{\mathcal{E}}(k_r, \omega, z)e^{i K z}$, reducing Eq.~\eqref{eq:uppe} to
\begin{equation} \label{eq:app-ode}
    \frac{\partial \tilde{\mathcal{E}}}{\partial z} = e^{-i K z} \tilde{S}\bigl(k_r, \omega, \tilde{\mathcal{E}}e^{iKz} \bigr) = e^{-i K z} \tilde{S}\bigl(\tilde{E}(z)\bigr),
\end{equation}
where the shorthand $\tilde{S}\bigl(\tilde{E}(z)\bigr)$ has been used for $\tilde{S}\bigl(k_r, \omega, \tilde{E}(z) \bigr)$. Equation~\eqref{eq:app-ode} can be integrated directly to obtain
\begin{equation}
    \tilde{\mathcal{E}}(z) = \tilde{\mathcal{E}}(z_0) + \int_{z_0}^z e^{-i K z'}\tilde{S}\bigl(\tilde{E}(z')\bigr)\,\mathrm{d}z'.
\end{equation}
Multiplying this solution by $e^{i K z}$ to again get $\tilde{E}$ and then evaluating at $z_0+\Delta z$ yields
\begin{align}
    \tilde{E}(z_0+\Delta z) &= \tilde{E}(z_0) e^{i K \Delta z} + \int_{z_0}^{z_0+\Delta z} e^{i K(z+\Delta z-z')}\tilde{S}\bigl(\tilde{E}(z')\bigr)\,\mathrm{d}z' \\
    &\approx \tilde{E}(z_0) e^{i K \Delta z} + \frac{\Delta z}{2}\left[ \tilde{S}\bigl(\tilde{E}^*(z_0+\Delta z)\bigr)+e^{i K \Delta z}\tilde{S}\bigl(\tilde{E}(z_0)\bigr) \right], \label{eq:app-soln}
\end{align}
where
\begin{equation} \label{eq:app-estar}
    \tilde{E}^*(z_0+\Delta z) = \tilde{E}(z_0) e^{i K \Delta z} + \frac{\Delta z}{2}\left(1 + e^{i K \Delta z} \right) \tilde{S}\bigl(\tilde{E}(z_0)\bigr).
\end{equation}
Equations~\eqref{eq:app-soln} and \eqref{eq:app-estar} give the solution.  This method of solving the UPPE is similar to Heun's method (a second-order Runge--Kutta method with two stages), and would be exactly Heun's method if $\tilde{E}^*(z_0+\Delta z) = \tilde{E}(z_0) e^{i K \Delta z} + \Delta z e^{i K \Delta z} \tilde{S}\bigl(\tilde{E}(z_0)\bigr)$ was used in place of Eq.~\eqref{eq:app-estar}.

\section{Parameter constraints} \label{sec:tanh}
The parameters contained in $\hat{A}(r)$, $\hat{\omega}(r)$, and $\hat{\tau}(r)$ can take any value determined by the optimization algorithm, including values that may not be physical.  For instance, the algorithm could adjust the bandwidth or pulse duration to be outside of the range achievable by a realistic laser system.  Additionally, it may be desired to limit the relative contrast in $\hat{A}(r)$, even while the energy is kept constant.  To constrain the values of the near-field functions, their raw output is passed through a hyperbolic tangent function before being used to compute the near-field amplitude $\hat{E}$.  As an example, let $\hat{\mathcal{A}}(r)$ denote the direct output of the coefficient function after parameterization with either B-splines or discrete points (see Sec.~\ref{sec:param}).  The coefficient function used in the calculation of Eq.~\eqref{eq:e-params} is then
\begin{equation}
    \hat{A}(r) = A_\mathrm{min} + \frac{1}{2}\left( A_\mathrm{max} - A_\mathrm{min} \right) \left\{ 1 + \tanh \left[3 \left( \hat{\mathcal{A}}(r) - 1 + A_\mathrm{shift} \right) \right] \right\},
\end{equation}
where $A_\mathrm{min}$ ($A_\mathrm{max}$) is the minimum (maximum) value allowed for the coefficient and $A_\mathrm{shift} = \log \left[(A_\mathrm{min} - A_0) / (A_0 - A_\mathrm{max}) \right] / 6$.  Defined in this way, initializing the raw output to $\hat{\mathcal{A}}(r) = 1$ corresponds to $\hat{A}(r) = A_0$, and any learned values $-\infty < \hat{\mathcal{A}} < \infty$ will yield an output $A_\mathrm{min} < \hat{A} < A_\mathrm{max}$.

\printbibliography

\end{document}